\documentclass[conference]{IEEEtran}
\IEEEoverridecommandlockouts

\usepackage{amsmath,amssymb,amsfonts}
\usepackage{graphicx}
\usepackage[caption=false]{subfig} % Use subfig instead of subcaption

\usepackage{cite}
\usepackage{hyperref}
\usepackage{balance}
\usepackage{flushend}
\usepackage{algorithm}
\usepackage{algpseudocode}
\usepackage{textcomp}
\usepackage{xcolor}
\usepackage{footnote}

\def\BibTeX{{\rm B\kern-.05em{\sc i\kern-.025em b}\kern-.08em
    T\kern-.1667em\lower.7ex\hbox{E}\kern-.125emX}}
\UseRawInputEncoding

\begin{document}

\title{Migration to Microservices: A Comparative Study of Decomposition Strategies and Analysis Metrics}

% And so on for additional authors

\author{\IEEEauthorblockN{ Meryam Chaieb}
 \IEEEauthorblockA{\textit{Laval University} \\
 \textit{Quebec, QC, Canada}\\
 meryam.chaieb.1@ulaval.ca}

 % \IEEEauthorblockN{ Khaled Sellami}
 % \IEEEauthorblockA{\textit{Laval University} \\
 % \textit{Quebec, QC, Canada}\\
 % khaled.sellami.1@ulaval.ca}
 \and
 \IEEEauthorblockN{ Mohamed Aymen Saied}
 \IEEEauthorblockA{\textit{Laval University} \\
 \textit{Quebec, QC, Canada}\\
mohamed-aymen.saied@ift.ulaval.ca}}
% \and
% \and
% \and
% \and
% \and

\maketitle
%##################################### Abstract#########################################
\begin{abstract}
The microservices architectural style is widely favored for its scalability, reusability, and easy maintainability, prompting increased adoption by developers. However, transitioning from a monolithic to a microservices-based architecture is intricate and costly. In response, we present a novel method utilizing clustering to identify potential microservices in a given monolithic application. Our approach employs a density-based clustering algorithm considering static analysis, structural, and semantic relationships between classes, ensuring a functionally and contextually coherent partitioning.
To assess the reliability of our microservice suggestion approach, we conducted an in-depth analysis of hyperparameter sensitivity and compared it with two established clustering algorithms. A comprehensive comparative analysis involved seven applications, evaluating against six baselines, utilizing a dataset of four open-source Java projects. Metrics assessed the quality of generated microservices.
Furthermore, we meticulously compared our suggested microservices with manually identified ones in three microservices-based applications. This comparison provided a nuanced understanding of our approach's efficacy and reliability. Our methodology demonstrated promising outcomes, showcasing remarkable effectiveness and commendable stability.
 
 Additional technical details are available in the replication packages \footnote{https://anonymous.4open.science/r/Migration-to-microservicess-B67F/README.md}
\newline Keywords-microservices architecture; static analysis; clustering; decomposition.
%Based on our findings, it can be concluded that our method outperforms the current state-of-the-art baselines across various metrics. Moreover, we also conducted a comparison of our method's resulting decomposition with those of two other popular clustering algorithms applied to similar problems. Our evaluation suggests that our solution is effective and stable, providing promising results.

\end{abstract}

%##################################### Introduction ########################################

\section{Introduction}\label{sec:1}
The monolithic architectures is one of the most widely utilized architectures for software design. In the realm of software architecture, the monolithic architecture stands as a prominent approach where an application is built as a single, indivisible unit. It encompasses all essential functionalities and components within a unified codebase, thereby presenting a tightly coupled system. This architectural style often involves a centralized database, user interface and business logic, rendering it self-contained and independent of external services. An exemplar of monolithic architecture, that we will use later in our evaluation process, can be observed in the context of the DayTrader \footnote{https://github.com/WASdev/sample.daytrader7 } application, a virtual stock trading platform. In this monolithic setup, all trading functionalities, user management, and financial calculations are contained within a single application. While this approach simplifies development and deployment and despite being used since the early days of software systems, it can pose challenges when it comes to scalability, maintaining code integrity and accommodating changes or updates in individual components \cite{app10175797} \cite{ benomar2015detection} \cite{vayghan2021kubernetes}. Monolithic architecture, in general, tends to experience performance issues when the amount of users exceeds a certain capacity level of these monolithic applications \cite{app10175797}.

Many methods have arisen throughout time to solve these performance difficulties, such as migrating to new technologies, managing independent services and deploying more powerful servers.
However, monolithic applications have evolved into massive, complex and often inefficient software systems over time, making them challenging to maintain. Additionally, they may not be able to support newer and more sophisticated technology \cite{r89} \cite{vayghan2018deploying} \cite{saied2018towards} \cite{huppe2017mining}\cite{saied2015visualization} \cite{ vayghan2019microservice}. Moreover, adapting these systems to meet rising user demand is either unfeasible or necessitates wasteful workarounds such as the duplication of the whole monolith \cite{newman2015microservices}.

Microservices architecture, in the other side, is gaining in popularity and is projected to play a large role in developing scalable, easy to maintain software products by focusing on tightly defined, separated services inside a distributed system \cite{app10175797}. The microservice architecture emerges as a contemporary approach where an application is built as a collection of small, independent services. These services are designed to be modular, self-contained and focused on specific business functionalities. Unlike the monolithic architecture, microservices operate as autonomous units that communicate with each other through well-defined APIs. This architectural style enables teams to develop, deploy and scale individual services independently, fostering flexibility and maintainability. For these reasons, numerous firms have sought to rework their monolithic apps into a microservice based version choosing a viable option by altering these systems while preserving the same functionality \cite{fritzsch_2019_migrationindustry} \cite{shatnawi2018identifying}. A noteworthy example of the microservice architecture can be found in the Netflix streaming platform. In this setup, various microservices handle distinct tasks such as user authentication, content recommendation, billing and media streaming. Each microservice can be developed, tested, deployed and scaled independently, allowing Netflix to rapidly innovate, adapt to changing demands and deliver a seamless streaming experience to its vast user base \cite{blinowski_monolithic_2022 }.

The transition from a monolithic design to a more durable and robust microservice architecture is based on the idea of finding contextually and functionally relevant modules and encapsulating them in a single service, while ensuring strong cohesion and low coupling between them.
As Rosati pointed out in their research on the migration cost \cite{rosati2019migrationcost}, transforming a mature monolithic software into microservices architecture may demand substantial investment in terms of time and cost. These difficulties have prompted academics to devise automatic decomposition methods that might ease the migration process. Such decomposition approaches seek to discover service boundaries and dependencies more efficiently and quickly, enabling for an easier transition \cite{r5}.
The task of transitioning a monolithic application into a microservices architecture is treated as a clustering problem in the context of our project. Our suggested method entails a multi-step procedure that employs static examination of the source code to determine the functional and contextual links between its classes. This stage is crucial for detecting relationships between classes and identifying potential service boundaries. Upon identifying the functional connections, we employ an adaptive density-based clustering technique known as adapted-BMSC to partition the classes into several prospective microservices. This selection is motivated by its superior performance in similar clustering tasks, outperforming other state-of-the-art algorithms. These resultant clusters represent potential microservices for further evaluation.
\newline We conducted an in-depth review utilizing a variety of metrics to measure the efficacy and efficiency of our method. We specifically evaluated the extracted microservices' quality to that of six other well-known decomposition baselines. To achieve a thorough review, we applied the evaluation criteria from several perspectives and evaluated them on seven example applications of diverse complexity.
\newline We conducted an extensive comparison with two widely-used clustering techniques that solve this challenge. We compared the performance of our technique to existing algorithms using various settings of hyperparameters. 
\newline On the other side , in addtition, we also  analyzed the efficiency of our technique by comparing the resultant microservices to those built by human specialists. This allowed us to assess how far our technique might automate the decomposition process while preserving the quality of human-designed microservices.
\newline As a culmination of these diligent efforts, our methodology has yielded a series of encouraging and promising outcomes as well as areas where more refinement and optimisation may be necessary.

The main contributions of our work are as follows:
\begin{enumerate}
    \item The proposed approach combines density-based clustering and static analysis techniques to leverage the advantages of both methods. It considers the structural and semantic dependencies among classes in a given monolithic application.
   
     \item A comparison between the resulting decomposition of the proposed algorithm and those of commonly used clustering algorithms in the field.

    \item A comparison between the microservices produced through our proposed approach and those that were manually identified by human experts.
\end{enumerate}

This paper is structured as follows:
Section \ref{sec2} presents the related work in the field of monolithic migration to microservices.
Section \ref{sec3} details the proposed methodology, including the clustering algorithms used. In Section \ref{sec4}, we discuss the findings of this effort and respond to different research questions.  Section \ref{sec5} outlines the threats to validity that were considered during the study. Finally, Section \ref{sec6}, concludes the work and discusses future research directions.

%%%%%%%%%%%%%%%%%%%%%%%%%%%%%%% RELATED WORK %%%%%%%%%%%%%%%%%%%%%%%%%%

\section{Related Work}\label{sec2}
Software engineering research has been tackling various issues in different phases of the software development lifecycle.
\newline  However, the fast pace of evolution in the IT industry  and the staggering growth of new technologies \cite{vayghan2021kubernetes}
based on APIs \cite{saied2015could1, shatnawi2018identifying, mujahid2021toward}, containers \cite{vayghan2019kubernetes},
microservices \cite{sellami2022improving, almarimi2019web, sellami2022hierarchical, saidani2019towards},
cloud and virtualization, put an increasing pressure on software development \cite{benomar2015detection} and deployment \cite{vayghan2019microservice, vayghan2018deploying}
practice to fully exploit this paradigm shift. This led to constant questioning of existing techniques \cite{saied2015could1} and results of software
engineering research \cite{saied2020towards, saied2018improving}, leading to investigating the use of AI and ML-based techniques to solve software engineering problems
in topics related to software reuse \cite{gallais2020api}, recommendation systems \cite{saied2016automated}, mining software repositories \cite{saied2020towards},
software data analytics and patterns mining \cite{saied2018towards, huppe2017mining, saied2016cooperative, saied2015mining} ,
program analysis and visualization \cite{saied2015observational, saied2015visualization}, testing in the cloud environment, Edge-Enabled systems \cite{mouine2022event},
microservices architecture \cite{sellami2022combining} and mobile applications.

The majority of methods in the literature that address the microservices extraction problem may be divided into two major components.
\newline The first component is concerned with the type of input provided to the solution and how it is handled. The methods suggested by MSExtractor \cite{sellami2022MSExtractor}, Bunch \cite{r7}, and \cite{r8} , for example, take as input the source code of a monolithic program and apply various static analysis techniques to it. MSExtractor and Bunch, in particular, construct call graphs that encode the relationships between the classes in these systems. On the other hand, the approach described in \cite{r8} turns the source code into a collection of Abstract Syntax Trees, which are then fed into a code embedding model \cite{aldebagy2021code2vec}. Static analysis, and more specifically the source code, is used with the assumption that structurally comparable classes or functions should be grouped together.

Some approaches have extracted the semantic relationships between the monolith's components from the source code. For example, the approach called HierDecomp \cite{r1} propose two types of measures : the structural similarity synthesised from the static calls between application's classes and the semantic similarity generated from the code text analysis.
Brito and al. \cite{r10} identify the systems' topics, based on topic modeling techniques, which correlate to domain terms and reflect the legacy system's microservices. The collection of lexical information in the source code, notably method declarations, variables, method and class names.., is used to infer such topic models.

Other approaches, such as Mono2Micro \cite{r2}, FoSCI \cite{r11}, and COGCN \cite{r12}, are based on the study of monolithic system use cases and execution traces. These solutions try to bring together classes or methods that interact at run-time for each business need given as input. For example, based on the execution traces, Mono2Micro computes similarity metrics across classes.

These analytic approaches are not mutually exclusive and can be used to provide improved results. CO-GCN \cite{r12}, for example, which, in addition to collecting execution traces, builds its model's architecture using the source code of the input application, assuming that each microservices contains classes with comparable domain concepts. Sellami and al \cite{sellami2022hydec} combine both static and dynamic analysis in order to cover the individual disadvantages of each of the analysis approaches.

There are, on either side, systems that employ different inputs, such as MEM \cite{r13}, which analyses the git commit history of monolithic programs. This technique generates a graph from the git history that encodes the class similarity.\\
Adding to that, Service Cutter \cite{r14} is a migration tool that uses the Unified Modeling Language (UML) diagrams to describe the various components of the monolithic application to be fragmented. The main drawback of this method is that the majority of existing programs lack representative diagrams, therefore it is up to the user to perform reverse engineering techniques to produce them and then transform them into the right format allowed by this approach. Nonetheless, Dehghan and al. \cite{r15} extended Service Cutter where the user will be required to provide the needed representative models as well as the source code to two distinct mechanisms: Service cutter \cite{r14} and MoDisco (Model Driven Reverse Engineering Framework) \cite{r16}.

The second component of each approach takes the data processed in the previous step as input and applies an algorithm to it in order to build the decomposition. Most methods utilize clustering algorithms, such as \cite{r8} which feeds vectors derived from code embedding into an Affinity propagation clustering process \cite{r17}. The similarity metrics computed by an agglomerative single-linkage clustering method \cite{r18} are used by Mono2Micro \cite{r2}. Based on the graph it developed, MEM \cite{r13} provides its own clustering mechanism. Based on the similarity metrics, HierDecomp \cite{r1} and HyDecomp \cite{sellami2022hydec} employ a DBSCAN \cite{r20} density based clustering algorithm which ends by having a hierarchical microservices decomposition recommendation.
Some methods suggest search algorithms to accomplish their goal. On the execution traces, MSExtractor \cite{saidani2019towards} use the non-dominated sorting genetic algorithm (NSGA-II) \cite{r19} whereas FoSCI \cite{r11} employs both NSGA-II and hierarchical clustering.
Bunch \cite{r7}, on the other hand, employs a hill-climbing algorithm.
A community discovery method is used by Service Cutter to provide a decomposition recommendation.

%###############################Appraoch###############################

\section{Proposed Approach}\label{sec3}

\begin{figure*}[ht]
    \begin{center}
        \includegraphics[width=\textwidth]{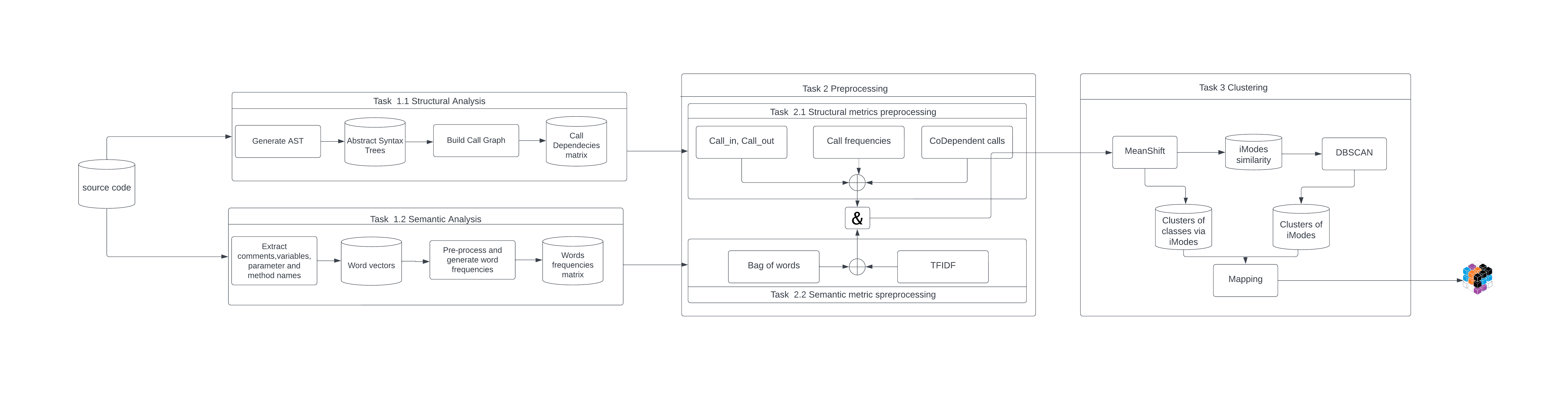}
    \end{center}
\caption{Overview of the Microservices Extraction Process.}
\label{fig:1}
\end{figure*}

%This section describes our suggested technique for discovering microservices candidates inside a given monolithic application.
The task of extracting microservices from a monolithic application is approached as a clustering problem, with the application's source code as input. Figure \ref{fig:1} depicts our approach in detail, outlining the phases involved in our research. Our primary goal in this effort is to achieve granularity at the class level.

Our methodology initiates by extracting both semantic and structural information from the application through static analysis of the legacy application's source code. This initial phase is followed by a preprocessing step where we systematically assess all possible combinations, opting for specific choices from both semantic and structural preprocessing components.
\newline Subsequently, the combined representations generated in the previous step are fed into the first clustering subtask, where we employ the Mean Shift algorithm. The centers of density identified through Mean Shift are then utilized to compute a novel distance metric termed "iModes similarity." This newly derived metric is subsequently fed into the final clustering subtask, which is executed by the Density-Based Spatial Clustering of Applications with Noise (DBSCAN) algorithm.
\newline The ultimate outcome of our approach is the resulting decomposition, achieved through the mapping of clusters initially detected by the mean shift algorithm and the final clusters generated by the DBSCAN algorithm. It is important to highlight that our methodology draws inspiration from the existing Boosted Mean Shift Clustering algorithm (BMSC), albeit with modifications to the calculation of iModes and adjustments in the input for the DBSCAN algorithm.

We have separated this section into two subsections to enhance clarity and comprehensibility, each presenting a unique viewpoint on our methodology. We will start by defining the problem we want to solve and outlining how we represent the monolithic system in the preprocessing step, as shown in Figure \ref{fig:1}. We will next move to the modeling step, where we will explain the primary clustering algorithms we employ.

\subsection{Representation of the Monolithic Application}
The input to our microservices extraction solution is a monolithic application, which is characterized as a set of Object Oriented Programming classes denoted as $C_M$=($c_1$, $c_2$,.., $c_N $), where N represents the total number of classes in the application. In this context, our approach aims to partition the original monolithic application into a set of K microservices, with the output being a collection of microservices, M = ($m_1$, $m_2$,.., $m_K $). Each microservice, $m_i$, represents a subset of the original classes and is defined as $m_i$=($c_a$, $c_b$,.., $c_p $), where $c_j$ is the OOP class that constitute the microservice. By applying our approach, we aim to optimize the decomposition of the monolithic application into microservices, where each microservice is expected to be cohesive and loosely coupled, resulting in a more maintainable and scalable architecture.

The initial  stage of our suggested solution focuses on representing the monolithic application and extracting the necessary information to build the microservices suggestions. To do this, we begin by creating an encoding scheme for each of the monolith's classes. The goal of our encoding approach is to capture the structural and semantic relationships that exist between the classes in the monolithic system  as described in the source code. We want to find important links between the classes by concentrating on these dependencies, which will allow us to produce more accurate and effective microservices suggestions. This encoding phase is crucial because it serves as the foundation for the following stages of our strategy, in which we use clustering techniques to discover groupings of similar classes that may be encapsulated within individual microservices.
\subsubsection{Structural encoding}
Abstract Syntax Trees (ASTs) can be created after the source code has been evaluated using a static analysis tool, such as "Understand" [9]. These ASTs help us comprehend the code structure and may be used to extract call relationships between classes. This is accomplished by creating call graphs that depict the relationships between distinct classes and how they interact with one another via function calls. We can establish which classes are commonly called together and uncover the most significant relationships in the codebase by studying these call graphs.% This knowledge may be used to break the monolithic application down into microservices.

As described in the diagram of figure \ref{fig:1} the structural information will be encodes using three different options: 

\begin{itemize}
    \item $Call_{in}$, $Call_{out}$: Each class in the monolithic application is encoded based on the dependency matrix derived from the static analysis phase. Specifically, we compute the sum of incoming and outgoing calls for each class. Our rationale for this encoding scheme is rooted in the observation that classes that exhibit frequent outgoing calls also tend to be called multiple times by other classes. By leveraging this insight, our approach seeks to group classes that interact frequently within the same cluster. This clustering strategy aims to reduce coupling between clusters while promoting greater cohesion within the resulting microservices
    \item Call frequencies: In contrast to the prior alternative, our second strategy tries to build more coherent clusters by encoding classes in greater depth. We analyse the frequency of calls between each pair of classes rather than just adding incoming and outgoing calls. By doing so, we hope to capture a more nuanced understanding of class connections, resulting in more coherent clusters.
    \item CoDependent calls: In this third and final structural encoding option, we take a more detailed approach to encode classes by considering the frequency of calls of classes that called both classes to encode each pair of classes, rather than just focusing on direct calls. This approach aims to group together classes that were involved in the same use case which will lead to produce even more cohesive clusters.
    \newline To aid in understanding this concept, we provide an example in Figure \ref{fig:2}.
\begin{figure}[htb]
    \begin{center}
        \includegraphics[width=\linewidth]{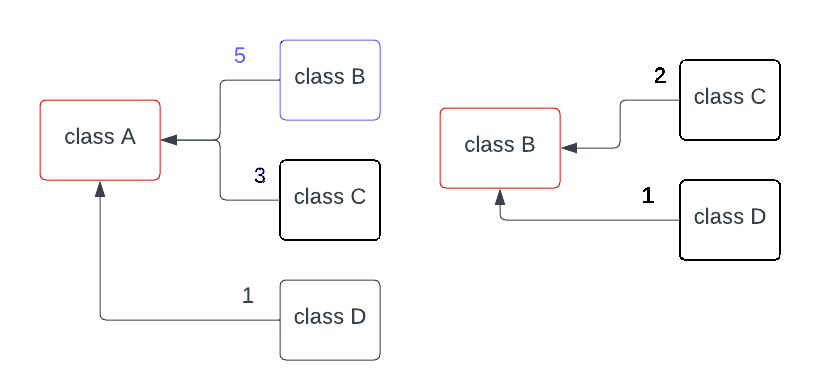}
    \end{center}
\caption{Illustrative example of CoDependent calls metric.}
\label{fig:2}
\end{figure}

In this scenario, there are four classes: A, B, C, and D. The objective is to encode the relationship between classes A and B. Figure \ref{fig:2} illustrates that class A is invoked five times by class B, three times by class C, and once by class D. In addition, class B is invoked twice by class C and once by class D. To encode the pair of classes A and B, the frequencies of calls originating from classes that invoked both A and B are summed. The encoding of the pair of classes A and B is the sum of incoming calls to A from the codependent classes C and D.
The basic idea behind this technique is that classes that are frequently called together are usually used to handle the same functionality, and hence are required for the same use case. As a result, the goal is to create robust microservices that can solve a specific use case each microservice.
    
\end{itemize}

\subsubsection{Semantic encoding}
Assume we are dealing with monolithic software projects that were created in accordance with industry norms. The names of classes, methods and variables are chosen based on functional principles in such projects and thorough annotations are included to indicate their intended use. As a result, the vocabulary employed in each software component can provide useful insights about the class's underlying domain,  meaning and functionalities.
It is critical to examine the semantic information associated with the classes when decomposing a monolithic program  into  microservices. This source of information   gives a more in-depth insight of the underlying concepts and class relationships in the legacy system. We can ensure that the resulting microservices are resilient and coherent by integrating this knowledge.

By including semantic information into the encoding process, we can determine the essential links between classes and the functionality they provide, facilitating the ability to combine them into coherent and self-contained microservices. This guarantees that each microservice serves a specific use case. Finally, this method results in a more modular and scalable system.

As a result, the semantic information of  each class is composed  of a collection of  terms that are used in different parts such as comments, parameter names, field names, method names, and variable names. To preprocess these words, we separate them using CamelCase, filter out stop words and normalise them using stemming. This method guarantees that we have a good comprehension of the class and its functionalities.

As seen in Figure \ref{fig:1}, the semantic information received from this preprocessing phase will be represented in two options:
\begin{itemize}
    \item Bag of Words (BoW) : One option for class encoding involves incorporating the frequencies of terms found within the vocabulary of the application. By doing so, we can ensure that the terms with higher frequencies are more closely related to the domain of the class. 
By considering the frequency of terms within each class, we can create a more refined understanding of the class and its intended purpose. This information can be used to group similar classes together into cohesive microservices, improving the overall organization and functionality of the resulting software system.

\item Term Frequency-Inverse Document Frequency (TF-IDF): Utilizing TF-IDF instead of simple Bag of Words can improve class clustering in a variety of ways.
To begin, TF-IDF considers not only the frequency of a word in a specific class, but also the inverse document frequency, which assesses how unique a term is to a class in comparison to the total corpus of classes. This means that unusual and unique terms that are exclusive to a class will have a larger weight in the TF-IDF calculation and will be more informative of the domain and purpose of the class.
Second, utilising TF-IDF can help limit the influence of common keywords that are not specific to any single class, such as other generic terms, which can distort clustering results when Bag of Words are implemented.

Overall, by encoding the classes using TF-IDF, the resultant feature vectors will be more representational of the classes' distinct properties, resulting in more accurate and effective clustering findings.
\end{itemize}

\subsection{Clustering algorithms}
The objective is to extract microservices by encoding classes structurally and semantically using different combinations of options. To achieve this, the Adapted Boosted Mean Shift Clustering (BMSC)\cite{r24} algorithm, along with other well-known clustering algorithms such as Density-Based Spatial Clustering of Applications with Noise (DBSCAN) \cite{r23} and Mean Shift \cite{r27}, is experimented with. The goal is to compare the abilities of adapted-BMSC against those of the other two algorithms. 
In the following section, we will provide a detailed description of each algorithm.
\subsubsection{DBSCAN algorithm}
DBSCAN is a clustering technique used in spatial databases to detect clusters and noise. The user must specify  two hyperparameters, Eps and MinPts. The method uses these parameters to arrange densely related points into a single cluster. One major  benefit of DBSCAN  is that the user doesn't need to define the number of clusters.  Alternatively, based on the data and the provided hyperparameters, the number of clusters can be arbitrary detected , leading in more accurate clusters \cite{r20}.\\
\textbf{Hyperparameters : }
\begin{itemize}
    \item \textbf{Eps ($\epsilon$) :}  refers to the radius of the neighbourhood surrounding the cluster's central point. This parameter determines the densest area in the data collection.
 
 \item \textbf{MinPts :}  This is the bare minimum of points required to build a cluster.
 \end{itemize}
The value of the Eps parameter in DBSCAN can effect the number of dense clusters and the number of identified clusters. A higher Eps value, in particular, might result in fewer dense clusters being detected, which can reduce the overall number of clusters identified by the method.
While implementing DBSCAN, it is important to avoid setting the MinPts parameter too low, as it is with the Eps parameter. A low MinPts value may cause the algorithm to generate an excessive number of less dense clusters. 

\begin{figure}[ht]
   \begin{center}
      \includegraphics[width=\linewidth]{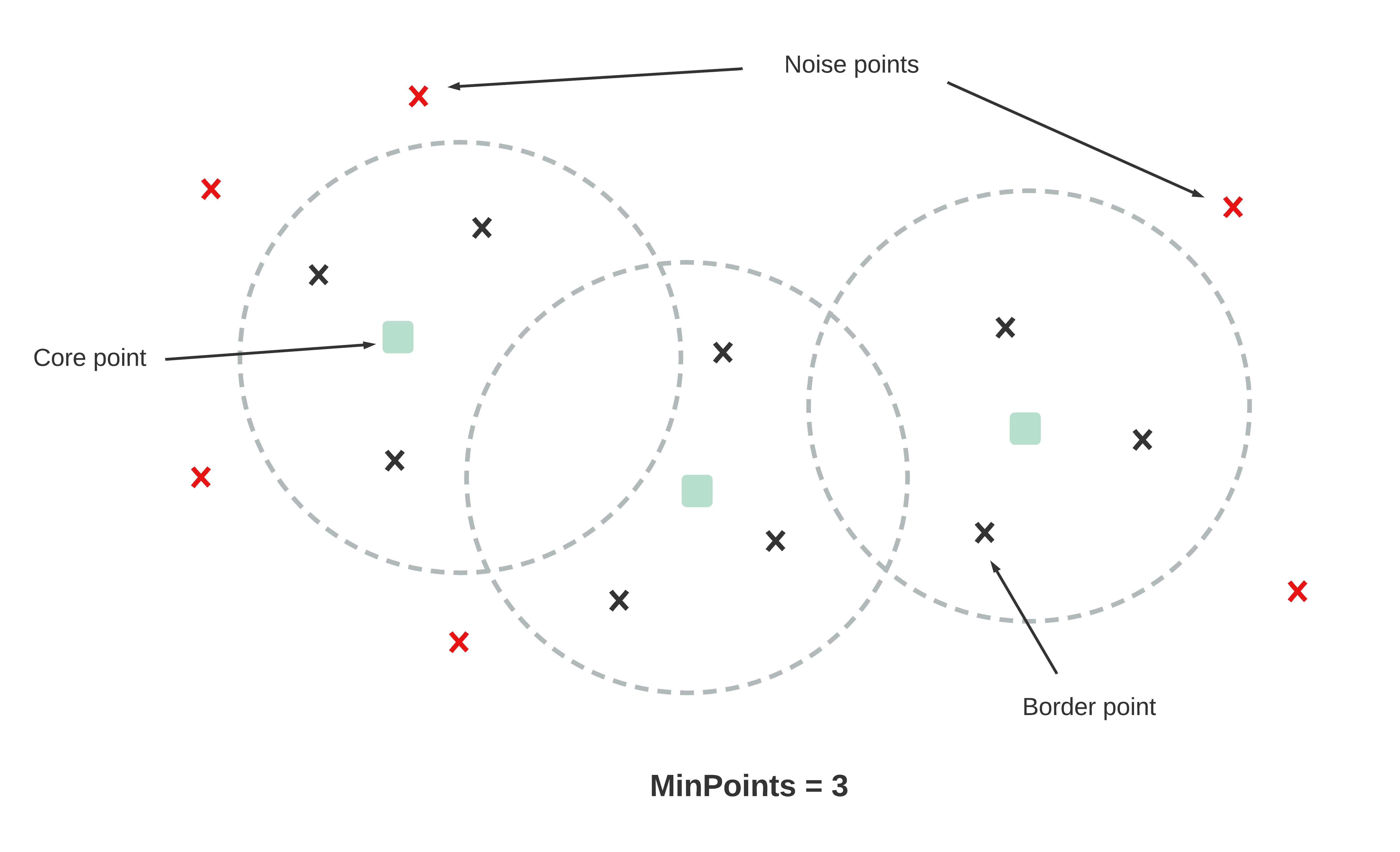}
 \end{center}
\caption{DBSCAN algorithm showcase.}
\label{fig:3}
\end{figure}

After executing the DBSCAN algorithm on a dataset, the results may be classified into three classes of points, as illustrated in Figure \ref{fig:3}. A core point is one that has at least MinPts number of points within an Eps radius. A Border point is any point that is close to Eps and possesses one or more Core points. Lastly, a Noise point is any point that is neither Core nor Boundary.

To build clusters, the DBSCAN algorithm goes through numerous phases. It begins by picking an arbitrary point in the database to serve as the first Core point. It then collects data points within a distance equal to Eps. A cluster is produced if the total number of points acquired is more than or equal to the minimum number of points necessary (MinPts). To enlarge the original cluster, this procedure is repeated for each cluster point. During this step, the algorithm creates the first cluster. The procedure is then repeated after  removing all of the points that composed it from the database. When no further clusters can be produced with the provided parameters, the algorithm stops. The rest of the points are labelled as Noise.

Despite the fact that DBSCAN is widely used in our problem, it still has limitations.
As demonstared in previous studies on extracting microservices from monolithic software relied on DBSCAN, this algorithm is highly sensitive to its hyperparameters, leading to significant variation in microservices' quality. Moreover, DBSCAN-based approaches may not work well with datasets with varying densities or non-globular shapes.

\subsubsection{Mean Shift}
The Mean Shift method is a cluster analysis technique that does not need any assumptions about the underlying distribution of the data. Based on the data, it can automatically detect non-linearly formed clusters and compute the number of clusters \cite{r27}.
%\textbf{Hyperparameter : }
%\begin{itemize}
 %   \item \textbf{Kernel bandwidth :}  The kernel bandwidth becomes the Mean Shift's primary hyperparameter. Its value, which is difficult to specify, might impact the performance of the Mean Shift.

 %\end{itemize}
 Figure \ref{fig:4} depicts a succession of steps taken by the Mean Shift algorithm to find clusters. It begins by identifying a region of interest, which is indicated by the red area in the picture. The centre of density or centre of mass for that region, shown by the blue point, is then calculated. The mean shift vector is then generated, and the centre of the area is shifted along the vector until it corresponds with the centre of mass point. This procedure is performed until convergence is reached.
Members of the same group are points that converge to the same center of mass.

Although the Mean Shift method has shown excellent results, \cite{r24} demonstrates that adapted-BMSC outperformed Mean Shift in a similar clustering problem with more stable clustering.

Given the difficulties involved with clustering algorithms in situations where there is no obvious separation between clusters or where the number of clusters is uncertain, we decided to investigate alternatives to standard techniques. We picked the Boosted Mean Shift Clustering (BMSC) technique since it has showed higher performance in similar situations, and we compared the results of this adapted clustering algorithm to those obtained using other algorithms, including some that employ DBSCAN.
%%%%%%%%%%%%%% pseudo code%%%%%%%%%%%%%%%
\begin{algorithm}[ht]
\caption{adapted-Boosted Mean Shift Clustering}\label{alg:cap}
\label{alg:bmsc}
\small
\begin{algorithmic}[1]

\Require {X, width, height, Eps.}

\Ensure {the final clustering results $cl\_final$.}

\State{  Initialize Grid( X,width,height)  

\Comment{Distribute X over G = width $\times$ height cells.}}

\State{iModes $\gets \emptyset $}
\State {counter $\gets$ 1 }
\While{$counter !=3$}

\For{j $\gets$   1 \textbf{to G}}

\State{newiModes $\gets$ MeanShift($cellData_i$)}

\State{iModes.Append(newiModes)}

\Comment{collect the iModes of each cell of the Grid}
\EndFor

\State{ConfidenceAssignement(Semantic\_similarity)}

\Comment{Assign confidence values to classes in each cell}\\

\For{j $\gets$ 1 \textbf{to G}}

\State{CollectedData $\gets$ CollectNeighborhoodData(j, neighborhood\_structure) $\cup$ $cellData_j$}

\State{$cellData_j$ $\gets$ WeightedSampling(CollectedData)}

\Comment{ update $cellData_j$}

\EndFor

\State{ cl\_iModes, numberOfClusters $\gets$ DBSCAN (iModes\_similarity, Eps)}

\Comment{cl\_iModes is the clustering results of the iModes}

\If{numberOfClusters == lastnumberOfClusters}
    \State {counter++}
    
\Else{ }
    \State {counter $\gets$ 1}
\EndIf

\EndWhile

\State{ cl\_final $\gets$ DataAssignement(X,cl\_iModes)}
\end{algorithmic}
\end{algorithm}
\subsubsection{Adapted-BMSC algorithm}
The adapted-Boosted Mean Shift Clustering algorithm is a hybrid clustering technique that combines two well-known clustering techniques: Mean Shift and DBSCAN. It is a density-based clustering methodology that overcomes some of the limitations of both approaches and can find clusters of any form and size with varied densities without the need for a predetermined number of
clusters.\cite{r24}

The adapted-BMSC first applies the Mean Shift algorithm on the dataset to generate a set of initial cluster centres. The centres of these clusters are used as input for the subsequent steps, which implements the DBSCAN algorithm. Adapted-BMSC selects a sample of the data that captures the skeleton of the clusters in order to properly identify the data's underlying structure. Essentially, the goal of adapted-BMSC is to overcome the limits of individual clustering algorithms by combining the capabilities of Mean Shift and DBSCAN, resulting in a more powerful and accurate clustering approach.

\begin{figure}[ht]
\centerline{\includegraphics [width=\linewidth] {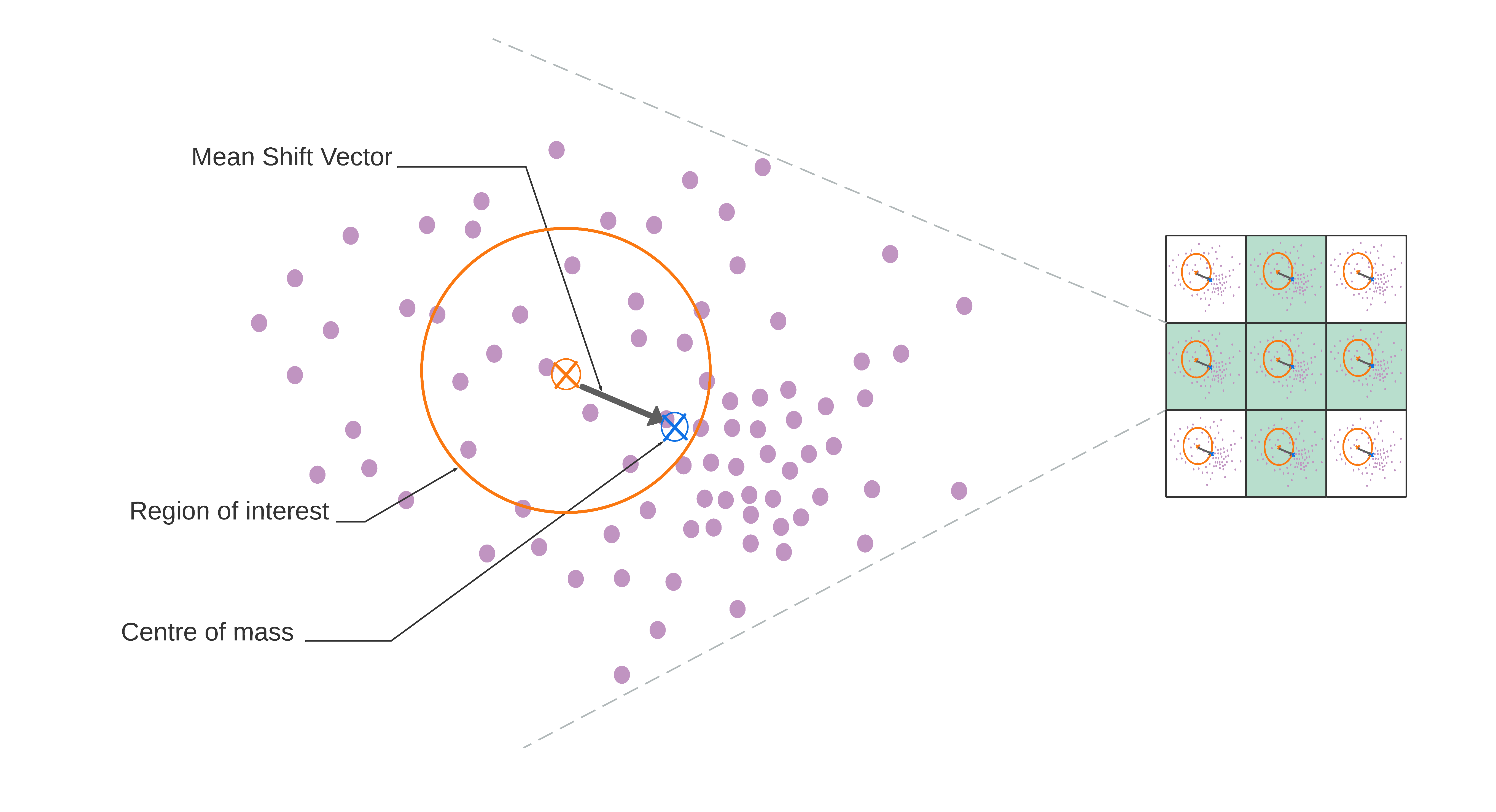}}
\caption{Mean Shift Algorithm showcase.}
\label{fig:4}
\end{figure}

\begin{figure}[ht]
\centerline{\includegraphics [width=\linewidth] {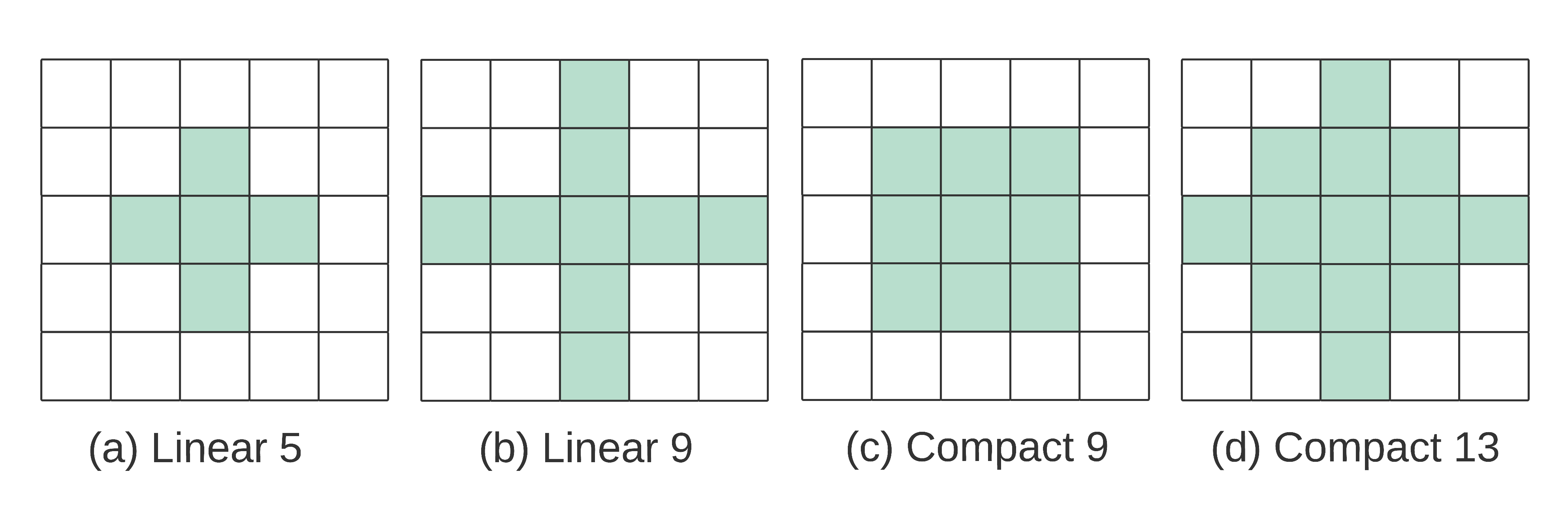}}
\caption{Potential neighbourhood structures.}
\label{fig:5}
\end{figure}

Algorithm \ref{alg:bmsc} outlines the steps involved in applying the adapted-BMSC algorithm. The first step is to divide the data uniformly into cells of a grid, where the grid size is specified by the user. Once the grid is initialized, the Mean Shift algorithm is applied independently to the data in each cell, as shown in Figure \ref{fig:4}. This produces a list of intermediate mode points (iModes) for each cell.
The next step in the adapted-BMSC algorithm is to disperse the data of each cell using a specific mechanism. This re-sampling mechanism involves each grid cell interacting with a limited number of cells in its vicinity, which are defined as its neighborhood based on a previously determined neighborhood structure. The BMSC paper \cite{r24} presents various neighborhood structures, which are depicted in Figure \ref{fig:5}. In our work, we adopt the linear 5 neighborhood structure.% , given the size of our grid.
%\vspace{-1cm}

Upon completing the initial step of re-sampling, the subsequent stage of the adapted-BMSC algorithm involves calculating the distances between all data points in the parent cell and those in its neighboring cells, relative to the intermediate modes (iModes) generated by the Mean Shift algorithm. To determine the similarity between the class and its corresponding iMode, we adopt a semantic similarity metric that assesses the confidence level of each relationship. By incorporating this metric, we can effectively identify the semantic association between the two points during this preliminary stage.
After that, adapted-BMSC algorithm utilizes the list of intermediate modes (iModes) obtained from the Mean Shift algorithm to run DBSCAN in order to identify clusters of densely packed iModes, which in turn generates clusters of the original data points at a lower level.
\newline In our particular scenario, we utilize an aggregation function to transform the iModes produced by the Mean Shift algorithm into a format similar to that of the legacy application's classes. More specifically, we represent each cluster center by summing the structural encodings of the classes assigned to its cluster, thus capturing the structural aspect of the mode point. Additionally, we compute the semantic part of the vector by summing the term frequencies of words used in those specific classes.

For the purpose of extracting reliable microservices, we adopt a novel approach inspired from the work of Sellami and al \cite{r1} where we don't directly input the encoders of iModes into the DBSCAN algorithm. Instead, we provide the connections between each pair of iModes. To achieve this, we employ the iModes similarity  measures that capture the structural and semantic relationships between the iModes. This approach aims to produce microservices that are consistent from both the implementation and use cases perspectives. 
\newline The iModes similarity is calculated as follows :

\begin{itemize}

\item\textbf{iModes Similarity (MS) :} It is a weighted sum of two similarity metrics, as provided by equation \ref{eq:imodesim}.
\end{itemize}

 \begin{equation}\label{eq:imodesim}
 \small
     MS(m_i,m_j)= \alpha  Sim_{str}(m_i,m_j) + \beta  Sim_{sem}(m_i,m_j)
 \end{equation}
 
With : 
\begin{itemize}
    \item \emph{$\alpha \in $ [0,1]}, 
    
    \item \emph{ $\beta \in $ [0,1]}, 
    
    \item \emph{$\alpha$ + $\beta$ = 1. }
\end{itemize}

Each one of the similarities is computed as follow:
\begin{itemize} 
    \item {\textbf{Structural similarity  ($Sim_{str}$)}} : We calculate this measure based on the number of method calls that are common between two iModes. This allows us to encode their level of interdependence and evaluate their similarity from a functional perspective.
    \newline The structural similarity of two given iModes $ m_i $ and $ m_j $ is determined using equation (2):
\end{itemize}
  
\begin{equation}
\resizebox{.9\hsize}{!}{
    $sim_{str}(m_i,m_j)=\left\lbrace 
    \begin{matrix}
        \frac{1}{2}{(\frac{call(m_i,m_j)}{call_{in}(m_j)} + \frac{call(m_i,m_j)}{call_{in}(m_i)} )} & If call_{in}(m_i)  \ne 0   and call_{in}(m_j) \ne 0\\
        \frac{call(m_i,m_j)}{call_{in}(m_j)} & If call_{in}(m_i) = 0 and call_{in}(m_j) \ne 0\\
        \frac{call(m_i,m_j)}{call_{in}(m_i)} & If call_{in}(m_i) \ne 0 and call_{in}(m_j) = 0	
    \end{matrix} \right. $
}
\end{equation}

With:
\begin{itemize}
    \item call( $m_i$ , $m_j$ ): refers to the number of times that method $m_i$ has called method $m_j$;
    \item $call_{in} $($m_i$): refers to the number of incoming calls in $m_i$.    
\end{itemize}

The structure similarity values range between 0 and 1, with 1 denoting that the iModes $m_i$ and $m_j$ are highly similar in functionality, and 0 indicating complete independence between the two.

\begin{itemize}
\item {\textbf{Semantic Similarity  ($Sim_{sem}$): }} In order to evaluate the similarity between the domain semantics of two iModes,% a metric is used which takes into account the preprocessing steps applied to prepare task 2.2 of the diagram in Figure \ref{fig:1}. To calculate the final vector, 
a TF-IDF model is used, as in equation \ref{eq:simsem}. The semantic similarity metric between two classes is represented by the cosine similarity between their respective vectors \cite{r22}.
\end{itemize}
\begin{equation} \label{eq:simsem}
\small
    sim_{sem}(m_i,m_j)=\frac{\vec{m_i}.\vec{m_j}}{||\vec{m_i}|| . ||\vec{m_j}||}
\end{equation}

With:
\begin{itemize}
    \item $\vec{m_i}$ :  represents the TF-IDF vector of iMode $m_i$.
    \item $|| \vec{m_i}||$ :  represents the Euclidean norm of the vector  $\vec{m_i}$.\\
\end{itemize}
The values of $ Sim_{sem}$ range from 0 to 1, with 1 indicating that both classes employ the same vocabulary and so fulfil the same use case.

Afterwards, we employ DBSCAN algorithm on the iModes similarity metric obtained from the previous step. The clustering process is iterated until DBSCAN algorithm produces the same number of clusters for three consecutive iterations.

Because of its unique combination of Mean Shift and DBSCAN, the adapted-BMSC method has outperformed conventional clustering approaches. This combination keeps both algorithms' non-parametric character, resulting in more robust clustering.\\

%In the upcoming section, we plan to investigate the research questions to gain a more comprehensive understanding of the benefits and drawbacks of each clustering algorithm and strategy. This analysis will enable us to identify the most efficient approach for this specific problem. 

%%%%%%%%%%%%%%%%%%%%%%%%%%%%%%%%% EVALUATION %%%%%%%%%%%%%%%%%%%%%%%%%%

\section{Evaluation}\label{sec4}
This section provides a summary of the evaluation process of our approach in discovering suitable microservices, which are available in the replication packages \footnote{https://anonymous.4open.science/r/Migration-to-microservicess-B67F/README.md}. Table \ref{tab:mono} summarizes the characteristics of the monolithic applications used to assess different aspects of our approach. %The section starts by introducing the research questions that the evaluation aims to answer, followed by an explanation of the evaluation metrics used to address these research questions. The study then moves on to investigate the most effective option among the many possibilities presented in the previous section. It then examines the hyperparameter sensitivity of various clustering algorithms. Lastly, the section presents the approach's findings and compares them to the outcomes of state-of-the-art approaches.
\begin{table}[ht]
\centering
\begin{tabular}{l c c c}
\hline  Project & Version & SLOC & \# of classes \\

\hline  Plants \footnote{https://github.com/WASdev/sample.mono-to-ms.pbw-monolith}  & 1.0 &  7,347 & 40 \\

DayTrader \footnote{https://github.com/WASdev/sample.daytrader7} &  1.4 & 18,224 & 118 \\

JPetStore \footnote{https://github.com/KimJongSung/jPetStore} &1.0& 3,341 &73\\
AcmeAir \footnote{https://github.com/acmeair/acmeair} &1.2 &8,899& 86\\
\hline

\end{tabular}
 \caption{- Characteristics of monolithic applications}
  \label{tab:mono}
\end{table}

%%% RQs
\subsection{Research Questions}
The goal of our experimental investigation is to address a set of research questions (RQs):
\newline \textbf{RQ1:} What is the most effective and promising configuration among the various choices in our approach that leads to favorable outcomes?
\newline \textbf{RQ2:} How does the stability and robustness of our approach compare to that of Mean Shift and DBSCAN in relation to hyperparameter variation?
\newline \textbf{RQ3:} How does our solutions perform in terms of partitioning quality when compared with state-of-the-art baselines?
\newline \textbf{RQ4:} How well do the extracted microservices compared to those that were manually identified by software engineers?

%%%% evaluation metrics
\subsection{Evaluation metrics}
We used a set of metrics specified in \cite{r2} to analyse various aspects of the extracted microservices without relying on the ground truth microservices:

\begin{itemize}
    \item {\textbf{Structural Modularity (SM) : } Determined by measuring the structural cohesiveness of classes inside a partition $m_i$ (scoh) and the coupling (scop) between partitions (M), as illustrated by equation \ref{eq:sm}.
    \begin{equation}\label{eq:sm}
    \small
    {SM = \frac{1}{M} }{ \sum_{i=1}^{M} scoh_i} -{ \frac{1}{(M(M-1))/2}}{  scop_{ij} }
    \end{equation}
    
    Where :
    \begin{itemize}
        \item \emph{$ scoh_i $ }= $ \frac{\mu_i}{m_i^2} $;
        
        \item \emph{$ scoh_i $}  = $ \frac{\mu_i}{m_i^2} $;
        
        \item \emph{ $\mu_i$} refers to the number of calls internal to the partition $m_i$;
        
        \item \emph{ $ scop_{ij} $ } =  $\frac{(\gamma_{ij})}{2*(m_i*m_j)}$;
        
        \item \emph{ $ \gamma_{ij}$} refers to the number of calls being made between partitions $m_i$ and $m_j$.
        
    \end{itemize}

    \textbf{ → The higher SM value, the better the decomposition.}
    }

     \item {\textbf{ICP :}  Depicts the percentage of calls that occur between two divisions as shown by equation \ref{eq:icp}.
    \begin{equation}\label{eq:icp}
    \small
    icp_{ij} = \frac{c_{ij}}{ \sum_{i,j=0 j/=i}^{M} c_{ij} }
    \end{equation}
    
    Where :
    
    \begin{itemize}
        \item  \emph{ $c_{ij}$} refers to the number of calls detected between partitions i and j.
    \end{itemize}

     \textbf{ → The lower the ICP value, the better the recommendation.}
    } 

    \item {\textbf{Interface Number (IFN) : } This metric, denoted as IFN, is used to count the number of interfaces present in a microservice $m_i$. An interface is defined as a class within $m_i$ that is invoked by a class within another microservice $m_j$. 
    \newline The calculation of IFN is described by Equation \ref{eq:ifn}.
    \begin{equation}\label{eq:ifn}
    \small
       IFN = {\frac{1}{N}} { \sum_{i=1}^{N} ifn_i } 
    \end{equation}

     Where :
     \begin{itemize}
         \item \emph{N}  refers to the total number of microservices;
         \item \emph{$ifn_i $} is the number of interface classes in the microservice $m_i$.
     \end{itemize}
     
     \textbf{ → The lower the IFN value, the better the recommendation.}
    }

        \item {\textbf{Non-Extreme Distribution (NED) : } This metric assesses the distribution of classes within microservices and aims to ensure that a microservice is neither too large nor too small. According to the study \cite{r2}, a microservice is considered non-extreme if it contains a number of classes within the range of [5, 20]. The metric is calculated using Equation \ref{eq:ned}.
    
    \begin{equation}\label{eq:ned}
    \small
       NED = 1 - {\frac{\sum_{k=0}^{N} n_k}{|N|}}
    \end{equation}

    Where :
    \begin{itemize}
        \item \emph{$n_k$} refers to the number of the non-extreme microservices;
        
        \item \emph{N} presents the total number of microservices.
    \end{itemize}
    
     \textbf{ → The lower the NED value, the better the recommendation.}
    }
\end{itemize}    

\subsection{Evaluation and Results for RQ1 }
%This subsection provides a detailed account of the evaluation protocol and presents the outcomes. It concludes by identifying the option that yielded the most effective results.

\subsubsection{Evaluation protocol}
The objective of this research question is to assess the quality of extracted microservices resulting from different combinations of structural and semantic information from classes within monolithic applications. The distinct strategies within our approach, elaborated in Figure \ref{fig:1}, are independently applied and yield diverse inputs for subsequent clustering tasks.
\newline To enhance comprehension of these combinations, we have assigned abbreviations to each one as follows:
\begin{itemize}
    \item Configuration 1 : $Call_{in}$,$Call_{out}$+ Bag of Words.
    \item Configuration 2 : $Call_{in}$,$Call_{out}$+ TFIDF
    \item Configuration 3 : Call Frequencies+ Bag of Words.
    \item Configuration 4 : Call Frequencies+ TFIDF.
    \item Configuration 5 : CoDependant calls+ Bag of Words.
    \item Configuration 6 : CoDependant calls+ TFIDF.
\end{itemize}

The aim is to contrast the outcomes of various Configurations and pinpoint the most efficient one in terms of representing the monolithic application for the decomposition task. As an initial phase of the evaluation, hyperparameters were set based on existing literature.
\begin{itemize}
    \item kernel bandwidth : Is set using the estimate bandwidth function from scikit-learn, which estimates the value of the bandwidth based on the provided data \cite{sklearn_api}.
    \item MinPts : Is set  to its default value ( minPts = 1 ).% set to 5 because a cluster is considered not extreme if its size ranges from 5 to 20 \cite{r2} for DBSCAN alone and\cite{r24}.
    \item Epsilon ($\epsilon$) : The critical hyperparameter $\epsilon$ is set using a k-distance graph \cite{sellami2022hierarchical}. % , where the distance to the k = minPts-1 nearest neighbor is plotted in descending order .
\end{itemize}
The DayTrader application is used to answer the first research question and its metadata is presented in Table \ref{tab:mono} for the purpose of evaluating and comparing the various strategies.

\subsubsection{Results}

% \begin{table}[ht]
% \centering
% \resizebox{\linewidth}{!}{
% \begin{tabular}{c c c c c c c }
% \hline  Metrics & Option 1 & Option 2 & Option 3 & Option 4 & Option 5 & Option 6 
% \\
% \hline    SM &   0.3696 & 0.3435 & 0.3887 & 0.4697 & 0.40545 & 0.4052

% \\
% \hline    IFN &    1.0344 & 1.250 & 1.0370 & 0.9677 & 1.0769 & 1.318

% \\
% \hline    ICP &  0.6500 & 0.591 & 0.618 & 0.6432 &  0.6257 & 0.639

% \\
% \hline    NED &     0.7241 & 0.6666 & 0.7037 & 0.7419 & 0.6538 & 0.636

% \\
% \hline    \# microservices &  29 &  24  & 27  & 31 & 26 & 22

% \\
% \hline    size of the largest micro  &  13 & 13 & 13 & 13 & 16 & 17

% \\

% \hline

% \end{tabular}}\\
%  \caption{Evaluation results of DayTrader Application. }
%   \label{tab:rq1}
% \end{table}
The presented Table \ref{tab:rq1} offers an overview of the evaluation results obtained from assessing the DayTrader Application across six distinct Configurations, each representing a unique combination of features.

Upon scrutinizing the diverse Configurations, the Structural Modularity metric (SM) values provide crucial insights into how effectively the generated microservices encapsulate the inherent structural relationships within the DayTrader Application. It is noteworthy that Configuration 4 significantly stands out with the highest SM value of 0.56. This prominence indicates that the fusion of Call Frequencies and TF-IDF in Configuration 4 yields microservices that adeptly capture the underlying structural dependencies. The elevated SM value indicates a strong alignment between Configuration 4's decomposition and the application's internal structure, suggesting the potential for more cohesive and organized microservices.

Shifting focus to the Interface Number (IFN) metric, which quantifies a microservice's interface dependencies, Configuration as long as Configuration 1 and 4 take the lead with the lowest IFN value of 0.93. This outcome implies that these strategies combinations effectively group features with fewer external interface dependencies. This attribute has the potential to foster the creation of self-contained and modular microservices.

Significantly, all configurations exhibit closely aligned Inter Call Percentage (ICP) values. Reducing communication between different components of the decomposition leads to a decrease in coupling, which can enhance the modularity and isolation of microservices, ultimately promoting component isolation.

Furthermore, the pivotal Non-Extreme Distribution (NED) metric aims to strike a balance between excessively large and excessively small microservices. Across the Configurations, NED values are closely clustered, signifying a favorable distribution that contributes to the overall quality of microservices' suggestion. Additionally, examining the number of detected microservices unveils variability among the Configurations, with Configuration 1 detecting the highest count (30) and Configuration 6 detecting the lowest count (22). A similar trend is apparent in the size of the largest microservice, with Configuration 6 featuring the largest (17) and Configuration 4 the smallest (12), both of which are within the classification recommandations outlined in the study \cite{r2}.

In essence, the intricate interplay of these metrics underscores the multifaceted nature of the microservices decomposition task. Each Configuration presents distinct strengths and trade-offs. The culmination of these metrics guides the selection of the most effective Configuration aligned with specific project goals and desired outcomes. Notably, Configurations utilizing TF-IDF in the semantic part showcase favorable evaluation values. Additionally, the utilization of CoDependent calls as a structural representation and TF-IDF as a semantic representation leads to the lowest NED and ICP metrics.

To conclude, the meticulous evaluation of these metrics showcases how various attributes influence microservices' quality. While each Configuration showcases noteworthy aspects, the harmony between structural coherence, interface independence, coupling reduction and a well-distributed size spectrum, as captured by NED, makes Configuration 6 a standout choice.
 
% Acording to the results presented in Table \ref{tab:rq1_1}, it can be concluded that option 6 is the best for the Mean Shift algorithm in terms of SM metric and performs similarly to the best options when comparing the other metrics. For DBSCAN, presented in Table \ref{tab:rq1_2}, option 6 has shown better results in terms of IFN, ICP, and NED, with a SM value that is close to being the best. Both DBSCAN and Mean Shift algorithms presented varied results, while BMSC had very similar results for all options and all metrics, presented in Table \ref{tab:rq1_3}. Furthermore, BMSC was able to detect a more stable number of microservices compared to the other algorithms, which often formed one large cluster or unique classes that did not meet the research goals. In contrast, the resultant microservices from BMSC were balanced and stable across different approaches, with the largest microservice containing a maximum of 17 classes as presented in Table \ref{tab:rq1} .
%\vspace{-1cm}
\begin{table}[ht]
\centering
\resizebox{\linewidth}{!}{
\begin{tabular}{c c c c c c c }
\hline  Metrics & Configuration 1 & Configuration 2 & Configuration 3 & Configuration 4 & Configuration 5 & Configuration 6 
\\
\hline    SM &   0.41 & 0.52 & 0.43 & 0.56 & 0.44 & 0.40

\\
\hline    IFN &    0.93 & 1.2 & 1.07 & 0.93 & 1.12 & 1.3

\\
\hline    ICP &  0.65 & 0.64 & 0.62 & 0.64 &  0.63 & 0.63

\\
\hline    NED &     0.73 & 0.67 & 0.69 & 0.75 & 0.67 & 0.63

\\
\hline    \# microservices &  30 &  25  & 26  & 32 & 25 & 22

\\
\hline    size of the largest micro  &  14 & 15 & 14 & 12 & 14 & 17

\\

\hline

\end{tabular}}\\
 \caption{Evaluation results of DayTrader Application. }
  \label{tab:rq1}
\end{table}

%%%%%% mean shift%%%%%%
% \begin{table}[ht]
% \centering
% \resizebox{\linewidth}{!}{
% \begin{tabular}{c c c c c c c }
% \hline  Metrics & Option 1 & Option 2 & Option 3 & Option 4 & Option 5 & Option 6 
% \\
% \hline    SM &  0.8526 & 0.7853 & 0.7944 &0.8614 & 0.8575& 0.8742

% \\
% \hline    IFN &   1.235 & 1.8  & 1.277 & 1.0454 & 1.0 & 1.214
% \\
% \hline    ICP &  1.0 & 0.9 & 1.0 & 1.0 & 1.0 & 1.0

% \\
% \hline    NED &  1.0  & 0.9 & 1.0 & 1.0 & 1.0 & 1.0

% \\
% \hline    \# microservices &  17 &  10  & 18  & 22 & 21 & 14

% \\
% \hline    size of the largest micro  &  98 & 102 & 97 & 92 & 97 & 104

% \\

% \hline

% \end{tabular}}\\
%  \caption{Evaluation results of DayTrader Application using Mean Shift algorithm. }
%   \label{tab:rq1_1}
% \end{table}

% \vspace{-0.5cm}

% %%%%%% DBSCAN %%%%%%

% \begin{table}[ht]
% \centering
% \resizebox{\linewidth}{!}{
% \begin{tabular}{c c c c c c c }
% \hline  Metrics & Option 1 & Option 2 & Option 3 & Option 4 & Option 5 & Option 6 
% \\
% \hline    SM &   0.120 & 0.1085 & 0.2702 &0.2718 & 0.2487 &0.1116

% \\
% \hline    IFN &   0.120 & 0.1085 & 0.2702 & 0.2718 & 0.2487 & 0.1116

% \\
% \hline    ICP &  0.3244 & 0.1426 & 0.1591 & 0.2859 & 0.3482 & 0.0079

% \\
% \hline    NED &   0.5 & 0.666 & 1.0 & 0.5 & 0.5 & 0.333

% \\
% \hline    \# microservices &  2 &  3  & 2  & 2 & 2 & 3

% \\
% \hline    size of the largest micro  &  108 & 86 & 116 & 113 & 113 & 104

% \\

% \hline

% \end{tabular}}\\
%  \caption{Evaluation results of DayTrader Application using DBSCAN algorithm. }
%   \label{tab:rq1_2}
% \end{table}

% \vspace{-0.5cm}
%%%%%% BMSC %%%%%%

%\vspace{0.5cm}
\fbox{
  \begin{minipage}{8cm} % adjust the width of the box as needed
    Configuration 6 proves to be the optimal approach, utilizing the CoDependent calls metric for structural information and TF-IDF vectors for semantic information. Consequently, our work will continue to focus on this strategy.
  \end{minipage}
}

\vspace{0.5cm}

%%%%%%%%%%%%%%RQ2%%%%%%%%%%%%%%%

\subsection{Evaluation and Results for RQ2 }
%In this part, we offer a detailed description of the evaluation procedure and report on the results. We examine the hyperparameter sensitivity of our approach using BMSC in comparison to DBSCAN and Mean Shift clustering algorithms.
\subsubsection{Evaluation protocol}
Within this protocol, our experimental design encompasses two primary phases, each serving a distinct purpose. The overarching objective is to meticulously examine the performance and gauge the sensitivity of the adapted-BMSC algorithm in comparison to the individual performances of DBSCAN and Mean Shift algorithms.

In the initial stage of our experimentation, we conducted individual tests for each algorithm. Subsequently, we embarked on a comprehensive exploration of hyperparameters, a pivotal facet in algorithmic performance. To ensure a thorough assessment, we systematically delineated the potential values for each hyperparameter. With a focus on rigorous control, we kept the other hyperparameters constant while varying a specific hyperparameter. For each conceivable value within the defined range, we executed the respective algorithms, recording the extracted microservices at each iteration. To comprehensively evaluate the results, we employed the suite of evaluation metrics, while concurrently plotting the metric values at each step to visualize their trends.
\newline Our investigation centered around the DayTrader monolithic project, a widely recognized benchmark within this domain.

To elucidate the hyperparameter exploration process, we specifically targeted two hyperparameters of significance:
\begin{itemize}
    \item \textbf{Kernel Bandwidth:} The estimate bandwidth function from the scikit-learn library informed our estimation of the central kernel bandwidth value. Subsequently, we systematically varied this parameter across a range that encompassed the estimated  value, allowing for a thorough assessment of its impact on algorithmic performance.
    \item \textbf{Epsilon ($\epsilon$):} With the aim of probing the influence of this hyperparameter, we systematically traversed the spectrum of Epsilon values, ranging from 0 to 1 with increments of 0.05. 
\end{itemize}

By meticulously investigating these hyperparameters, our protocol endeavors to unravel the intricate dynamics of algorithmic performance and sensitivity, contributing to a nuanced understanding of adapted-BMSC, DBSCAN and Mean Shift within the context of software clustering.

\subsubsection{Results}
In our initial analysis experimentation, the comparison results among the three algorithms are illustrated in Table \ref{tab:rq2} where the results were detected using the estimated bandwith and the epsilon using K- distance Graph while keeping the MinPts set to 1. This comparison of clustering algorithms' evaluation outcomes offers captivating insights into their performance within the context of decomposing monolithic applications.Remarkably, Mean Shift exhibits a notably high structural modularity (SM) value of 0.87, signifying its proficiency in capturing structural relationships. Nevertheless, a more nuanced evaluation of its suitability for the task is warranted. Interestingly, despite adapted-BMSC displaying a lower SM value of 0.53, which is similar to the DBSCAN value of 0.54, further explanation and observations reveal adapted-BMSC as the leading algorithm due to its highly effective decomposition outcomes.

Turning our attention to Interface Number (IFN), DBSCAN boasts the lowest value at 0.34, suggesting its adeptness in generating microservices with fewer dependencies on external interfaces, a pivotal trait for effective modularization.

The assessment of Inter-Call Percentage (ICP) emphasizes the significance of reducing coupling. All approaches, namely DBSCAN, adapted-BMSC, and Mean Shift, maintain ICP values of 0.65, 0.63, and 0.61, respectively, demonstrating their commitment to minimizing communication between distinct components.

An illuminating aspect arises when delving into the number of microservices and the size of the largest microservice. Adapted-BMSC stands out with the detection of 22 microservices, showcasing its proficiency in skillfully segmenting the application into manageable components. This count starkly contrasts with DBSCAN's 88 and Mean Shift's 14 microservices. Additionally, adapted-BMSC demonstrates a well-balanced size distribution among its largest microservices, each with a size of 17, in stark contrast to Mean Shift and DBSCAN, where the largest microservices encompass 104 classes for Mean Shift and 9 for DBSCAN. This balance notably enhances the overall quality of the decomposition process.

In essence, adapted-BMSC's success in achieving the fundamental objectives of microservices decomposition, despite its seemingly lower structural modularity (SM) value compared to the results of other algorithms, underscores its holistic approach. By effectively tackling inherent issues in monolithic applications and attaining optimal interface independence, coupling reduction, and balanced distribution, adapted-BMSC emerges as the preferred algorithm for this specific decomposition task.

\begin{table}[ht]
\centering
\resizebox{\linewidth}{!}{
\begin{tabular}{c c c c  }
\hline  Metrics & Mean Shift & DBSCAN & adapted-BMSC 
\\
\hline    SM &   0.87 & 0.54 & 0.53 

\\
\hline    IFN &    1.21 & 0.34 & 1.03

\\
\hline    ICP &  0.61 & 0.65 & 0.63

\\
\hline    NED &     1 & 0.97 & 0.72

\\
\hline    \# microservices &  14 &  88  & 29  

\\
\hline    size of the largest micro  &  104 & 9 & 16 

\\

\hline

\end{tabular}}\\
 \caption{Evaluation results comparison of clustering algorithms using DayTrader application. }
  \label{tab:rq2}
\end{table}

\begin{figure*}[ht]
\centerline{\includegraphics [width=\textwidth] {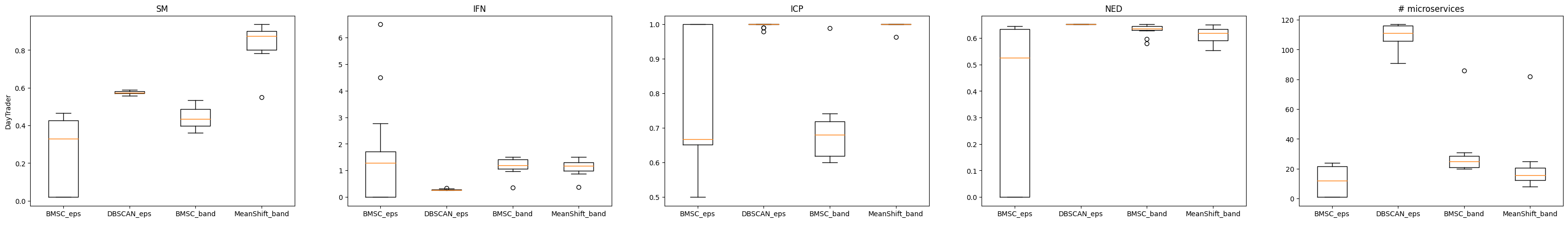}}
\caption{Evaluation metrics for different hyperparameters values when extracting microservices from the project DayTrader.}
\label{fig:6}
\end{figure*}

Transitioning to the second phase of analysis, we delve into the examination of hyperparameter sensitivity. The insights garnered from Figure \ref{fig:6} provide clear evidence that adapted-BMSC exhibits higher sensitivity compared to DBSCAN when subjected to variations in hyperparameters. Specifically, when assessing the epsilon hyperparameter ($BMSC_{eps}$ vs. $DBSCAN_{eps}$) and the bandwidth hyperparameter ($BMSC_{band}$ vs. $Mean\ shift_{band}$) across all metrics, adapted-BMSC's responsiveness is notably more pronounced.

Despite DBSCAN's superior performance in terms of SM and IFN metrics in comparison to adapted-BMSC, it yields a substantial number of microservices, averaging around 115 for an application containing 118 classes. Regrettably, this outcome does not align with our objectives. In contrast, adapted-BMSC demonstrates a more tempered sensitivity when varying the bandwidth hyperparameter compared to its sensitivity when altering the epsilon hyperparameter ($BMSC_{band}$ vs. $BMSC_{eps}$). This distinction arises from the fact that varying the bandwidth has the potential to generate differing numbers of modes, which are subsequently interconnected via DBSCAN. In contrast, changes in the epsilon hyperparameter directly influence the ultimate count of microservices, a fact underscored by the observed variations in the number of microservices.

\vspace{0.5cm}
\fbox{
  \begin{minipage}{8cm} % adjust the width of the box as needed
     In contrast to the findings in \cite{r24}, our analysis suggests that for our case, adapted-BMSC is more susceptible to the selection of its hyperparameters, specifically the epsilon hyperparameter, compared to DBSCAN and Mean Shift when used independently. However, adapted-BMSC demonstrates greater consistency in the resulting decomposition, as evidenced by its better NED values compared to other algorithms where the resulting microservices are either too large or too small.
  \end{minipage}
}

\vspace{0.5cm}

%%%%%%%%%%%%%%%%%%%%%%%%%%%%%%%%%%%%%%%%%%%%%%%%%%%%%%%%%%%%%%%%%%%%%%%%%%%%%%

\subsection{Evaluation and Results for RQ3 }
%This subsection outlines the evaluation protocol, presents the obtained outcomes, and compares them with the state-of-the-art baseline methods to determine the effectiveness of our approach.

\subsubsection{Evaluation protocol}
In order to evaluate the quality of our proposed solution, we compared it with six existing baselines include Bunch\cite{r7}, CoGCN\cite{r12}, FoSCI\cite{r11}, MEM\cite{r13}, Mono2Micro\cite{r2}, and HierDecomp\cite{r1}. For this purpose and for a comprehensive evaluation, we experimented with all the approaches on four monolithic applications with varying complexity, namely DayTrader, Plants, JPetStore and Acmeair as presented in Table \ref{tab:mono}.

To account for the hyper-parameter sensitivity of each solution, we generated multiple microservices decompositions from each baseline by varying their corresponding hyperparameters. 

\subsubsection{Results}

 \begin{figure*}[ht]
\centerline{\includegraphics [width=\textwidth] {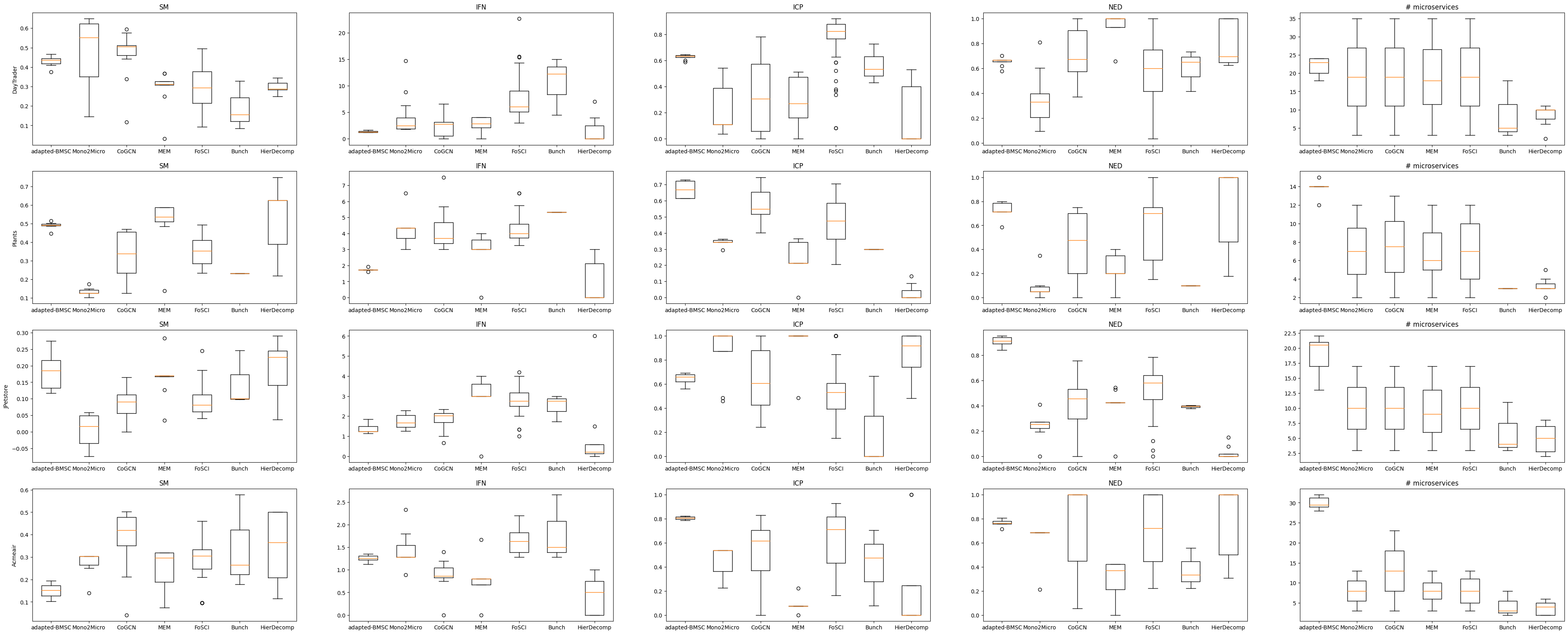}}
\caption{Boxplot Analysis of Project/Baseline/Metric Combinations.}
\label{fig:7}
\end{figure*}

Figure \ref{fig:7} provides a visual representation of the results through boxplots, offering a comprehensive view of the outcomes. Each row corresponds to a distinct project, and each column represents different metrics, considering various methodologies. In particular, let's focus on the first column, which pertains to the structural modularity metric (SM). In this context, our method stands out as a strong contender. It exhibits significant superiority over MEM, FOSCI, Bunch and HierDecomp when applied to the DayTrader project. However, when considering the Plants project, our method lags slightly behind HierDecomp and MEM, with a marginal difference of 0.15 in their mean values, while outperforming all other baseline methods.When we turn our focus to the last two projects, our solution consistently demonstrates superior performance compared to Mono2Micro, CoGCN, MEM, FOSCI, and Bunch when examining the JPetstore project. However, it's worth noting that the hierdecomp approach achieves the highest structural modularity value at 0.6. As for the acmeair project, we attained the lowest mean SM, with CoGCN securing the top rank.

Shifting our focus to the second column, which addresses the Interface Number (IFN) metric, our approach proves its effectiveness. Across the DayTrader, Plants, and JPetStore projects, it consistently maintains the lowest mean value compared to all baseline methods except Hierdecomp. In the Acmeair project, we outperformed Mono2Micro, FOSCI, and Bunch. This further emphasizes the robust performance of our method in efficiently managing interface numbers across a diverse range of projects.

Transitioning to the fourth column, which corresponds to the Non-Extreme Distribution (NED) metric, our approach demonstrates balanced performance. We outperformed hierdecomp's results in all projects except for JPetStore. When compared to Bunch and MEM, our results are notably favorable, closely resembling those obtained with CoGCN, especially in the case of the DayTrader application. Additionally, concerning the Acmeair project, our approach outperforms CoGCN and aligns closely with FOSCI. However, for the JPetStore application, HierDecomp records the best NED values.

It is worth noting that our approach does exhibit a slightly higher Inter-Call Percentage (ICP) value compared to most baseline methods. The exceptions are FOSCI, which records the highest ICP value for the DayTrader project, and HierDecomp, Mono2micro, MEM, as well as an outlier result from FOSCI for the JPetStore application.

Examining the fifth and final column, which pertains to the number of generated microservices, our approach consistently demonstrates robustness in its decomposition results, displaying minimal variability when hyperparameters are adjusted. The only exception to this pattern is a narrow range of variability observed in the HierDecomp approach. Notably, our approach often leads to a higher number of microservices compared to all other approaches, as is particularly evident in the AcmeAir and JPetStore projects.

One notable advantage of our approach is its remarkable stability, especially in terms of the number of identified microservices. This steadiness emphasizes its robustness and underscores its efficacy when contrasted with alternative methodologies.

\vspace{0.5cm}
\fbox{
  \begin{minipage}{8cm} % adjust the width of the box as needed
     The comparison results suggest that our proposed solution yields promising outcomes when compared to the baseline approaches, particularly in terms of SM, IFN, and NED metrics. However, this enhancement is associated with higher values in the ICP metric.
  \end{minipage}
}

%%%%%%%%%%%%%%%%%%%%%%%%%%%%%%%%%%%%%%%%%%%%%%%%%%%%%%%%%%%%%%%%%%%%%%%%%%%%%%
\subsection{Evaluation and Results for RQ4 }
\subsubsection{Evaluation protocol}

To address RQ4, we opted to examine three Java-based projects that utilize microservices architecture, each exhibiting varying levels of complexity. These projects are presented in Table \ref{tab:micro}
%table of apps
\begin{table}[ht]
\centering
\resizebox{\linewidth}{!}{
\begin{tabular}{l l l l l}
\hline  Project & Version & SLOC & \# of classes  & \# of microservices\\

  \hline Spring PetClinic \footnote{https://github.com/spring-petclinic/spring-petclinic-microservices} &  2.3.6 &  1,889 &  43 & 7\\

 Microservices Event Sourcing \footnote{https://github.com/chaokunyang/microservices-event-sourcing}  & 2.8.0  &  4,597 &  121 & 12\\

 Kanban Board \footnote{https://github.com/eventuate-examples/es-kanban-board} & 0.1.0  & 4,380 &  118 & 21\\
\hline

\end{tabular}}
 \caption{- Characteristics of Microservice-based applications}
  \label{tab:micro}
\end{table}

In our research study aimed at evaluating the effectiveness of our proposed approach, it is crucial to define appropriate metrics that can accurately compare two sets of classes representing the extracted microservices with their corresponding ground truth microservices. However, the process of identifying the corresponding ground truth microservice for each extracted microservice is a challenging task. To overcome this challenge, inspired from the work of Sellami and al \cite{r1},we have developed a method that utilizes the number of common classes between the extracted microservice and each of the ground truth microservices to determine the corresponding microservice.

To be specific, we introduced a function represented by equation \ref{eq:corr} that takes an extracted microservice $m_i$ and a set of ground truth microservices M as input and selects the ground truth microservice with the highest number of common classes with the extracted microservice.
\begin{equation}\label{eq:corr}
\small
       Corr (m_i , M) = argmax_{m_j \in M }( \frac{| m_i \cap m_j|}{|m_i|})
\end{equation}

After identifying the ground truth microservices that correspond to each extracted microservice using our suggested method, it is necessary to calculate relevant metrics in order to evaluate our methodology. Among the statistics used is precision, which is calculated using equation \ref{eq:prec}.

\begin{equation}\label{eq:prec}
\small
       Precision = \frac{1}{|M|}\times \sum_{\forall m_i \in M} \frac{|m_i \cap Corr(m_i,M_t)|}{|m_i|}
\end{equation}

Precision is a measure of how accurate our technique is in identifying the classes that belong to each microservice. It provides the average proportion of correctly recognised classes relative to the total number of identified classes for each extracted microservice. Through the calculation of precision, we can gauge the effectiveness of our approach in accurately identifying the classes that correspond to each microservice.

To evaluate the effectiveness of our approach in identifying the microservices, we also compute another metric called the Success Rate (SR). The SR is calculated using equation \ref{eq:SR} and measures the percentage of successfully retrieved microservices based on the precision metric.
\begin{equation}\label{eq:SR}
\small
       SR = \frac{1}{|M|}\times \sum_{\forall m_i \in M}matching(m_i,Corr(m_i,M_t))
\end{equation}

The SR provides a complementary perspective to the precision metric in assessing the performance of our approach. It takes into account the overall number of correctly identified microservices and their precision, providing a more comprehensive evaluation of our approach's effectiveness where :
\begin{equation}
\small
\resizebox{.9\hsize}{!}{
   $ matcing(m_1,m_2)=\left\lbrace 
    \begin{matrix}
         1 & if \frac{|m_1 \cap m_2|}{|m_1|} \geq threshold \\
         0 & otherwise	
    \end{matrix} \right. $
}
\end{equation}
Specifically, we consider two sets of classes, $m_1$ and $m_2$, and a threshold value, threshold $\in [0, 1]$. Using these inputs, we calculate the success rate (SR) at k, for a given k value ranging from 1 to 10.

In order to evaluate the performance of our microservices decomposition approach, we take the following steps. Firstly, we collect all Java classes for each test project and combine them to form a Monolithic architecture. This serves as the input to our approach, while the original version of the project is considered as the ground truth decomposition for comparison. Next, we generate multiple microservices decompositions using different hyperparameter values, and for each decomposition, we calculate several metrics such as precision, SR@5, SR@7, and SR@9, along with the corresponding true decomposition. This enables us to analyze the effectiveness of our approach in generating microservices decompositions for a wide range of hyperparameter values.

\subsubsection{Results}
\begin{figure*}[ht]
\centerline{\includegraphics [width=\textwidth] {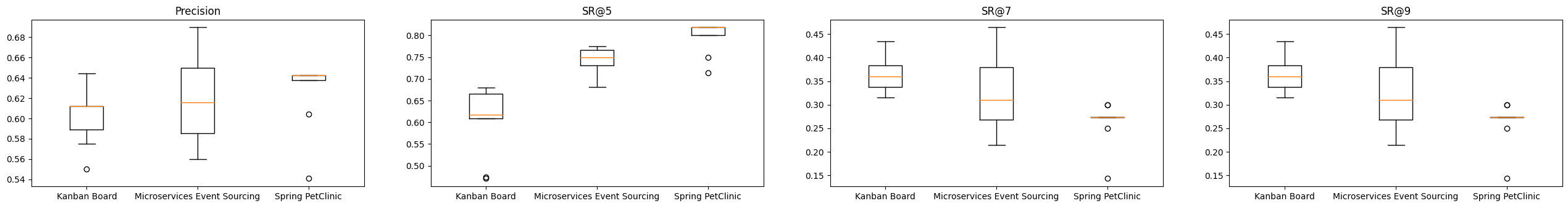}}
\caption{A comparative analysis of generated vs true decompositions through boxplots.}
\label{fig:8}
\end{figure*}

The study's results are shown in Figure \ref{fig:8} through boxplots for each project and metric. Each project's median precision values are within the range of 0.6 to 0.65. More specifically, the Kanban Board demo, Microservices Event Sourcing projects and spring petclinic project achieved precision values greater than 0.56 for all decompositions except for one outlier in the Kanban Board application adn for the spring petclinic with a precision value close to 0.55. The precision variability was reasonable, with a maximum score variation goes for the Microservices Event sourcing application and the smallest variability observed in the spring petclinic project. These results suggest that the method remains stable despite the increase in the number of classes, but projects with a smaller number of classes perform better.

As for the success rate, we observed that as the threshold increases, the median and maximum scores for each project drop. However, the Kanban Board demo and Microservices Event Sourcing projects have less variance in their values. At SR@5, all projects achieved a precision higher than 0.6 for all decompositions, except for an outlier close to 0.5 detected in the Kanban Board project. The spring petclinic project had the highest precision at this rate, exceeding 0.8 as a median. Notably, the scores for SR@7 to the strictest SR@9 were the same for all projects. These results suggest that a high percentage of microservices in all decompositions achieved a median precision score higher than 0.25. Moreover, as the project's size decreased, the variance of the results pattern decreased.

The variability observed in the derived microservices outcomes of the Microservices Event Sourcing project could potentially be attributed to its incorporation of multiple natural languages for domain terms, a distinction not present in the other projects that solely employ English. This linguistic diversity poses a challenge to the re-sampling process, which might lead to misclassification of certain classes with their nearest iModes, thus impacting their accurate assignment to the appropriate microservices. Nonetheless, the outcomes attained by the Microservices Event Sourcing project remain in line with those of its counterparts. It's noteworthy that the Microservices Event Sourcing project exhibited superior performance, surpassing other projects with its highest precision value in the most stringent thresholds.

%figure

\begin{figure*}[ht]
\centerline{\includegraphics [width=\textwidth] {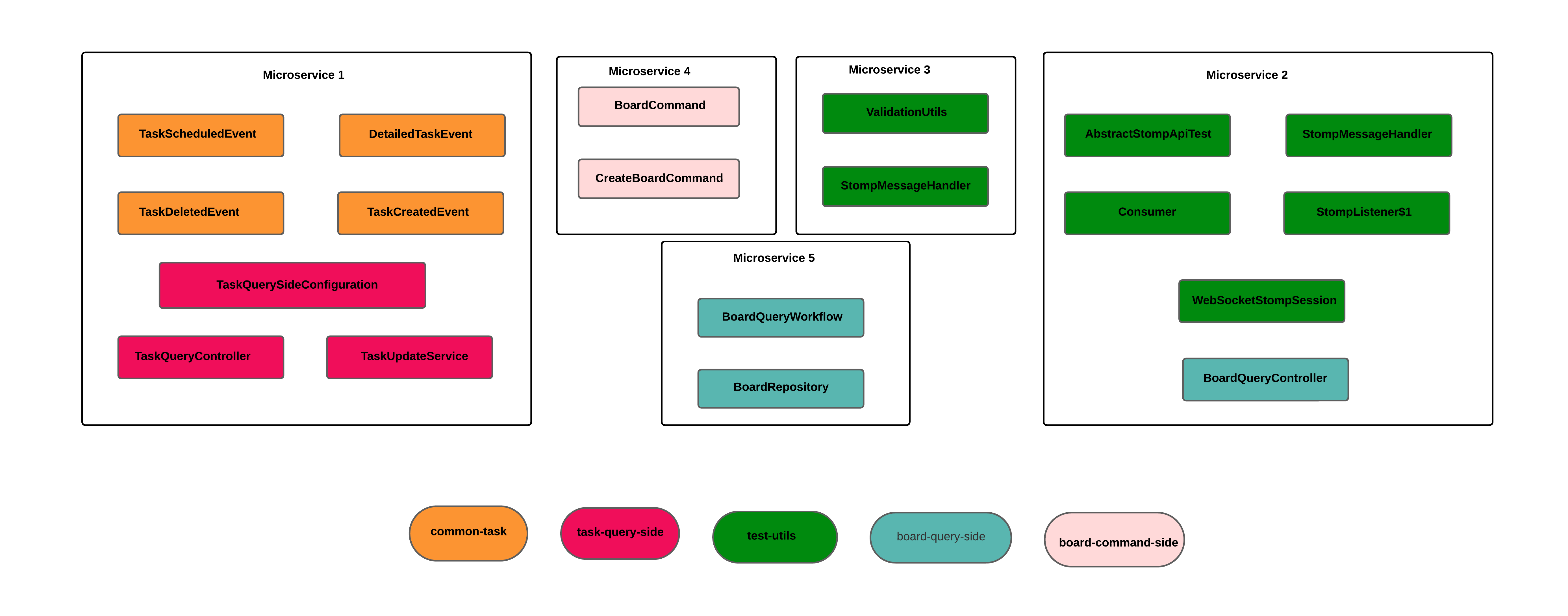}}
\caption{A subset of the microservices obtained from a decomposition of the project Kanban Board Demo.}
\label{fig:final}
\end{figure*}

For a more comprehensive analysis, we delve into the insights presented in Figure \ref{fig:final}. As a case in point, we examine the decomposition results of the Kanban Board demo project. This Java-based application, developed using Spring Boot, serves as a practical illustration of how the Eventuate™ Platform can facilitate the construction of real-time, multi-user collaborative applications. Specifically, the Kanban Board application showcases the collaborative creation and editing of Kanban boards and tasks. Any modifications initiated by one user on a board or task are instantly reflected to other users who are concurrently accessing the same board or task.

The architecture of the Kanban Board application relies on Eventuate™'s Event Sourcing-based programming model, which is optimally suited for such use cases. The application persists business objects such as Boards and Tasks, as a sequence of events that alter their state. Upon a user's action to create or update a board or task, the application records an event in the event store. Subsequently, this event is conveyed to subscribers with an interest in the event. Within the Kanban application, an event subscriber transforms each event into WebSocket messages, facilitating real-time updates in each user's browser interface.

Given the scale of the project, Figure \ref{fig:final} presents a subset of the microservices that resulted from one of the decomposition processes. In this illustrative representation, ellipses denote the original microservices' names, while large white rectangles symbolize the new microservices that have been generated. These rectangles encompass the classes, which are color-coded based on their originating microservices.

Importantly, it's pertinent to highlight that in this specific decomposition, the exact count of microservices matches that of the original application. However, it's essential to recognize that obtaining an identical distribution of classes is not necessarily guaranteed. A closer examination of the results reveals intriguing nuances. For instance, within Microservice 2, we observe an aggregation of 7 classes that were initially divided—4 residing in the task-query-side microservice and 3 in the common-task microservice. Our approach amalgamated these classes due to their shared implementation of task specifications and services.

"Let's explore the original Test-utils microservices in more detail. In our approach, its classes underwent partitioning into two distinct microservices, namely Microservice 2 and Microservice 3. Additionally, Microservice 4 emerged exclusively to encapsulate the board-command-side classes. Notably, the concept of Board-query-side also underwent partitioning, resulting in Microservices 5 and 2.

However, an interesting observation lies in the amalgamation of the BoardQueryController class with the test utils concept in our approach. If we focus more on this classification of classes, we find that the BoardQueryController class shares a significant number of words in its vocabulary with the remaining classes associated with the same microservice in our approach. Such words include "native," "code," "id," "notification," "class," "hash," "equals," "wait," "clone," and more. This intricate differentiation highlights the nuanced decisions our approach makes to optimize microservice composition while maintaining functionality and cohesion."

%figure

\vspace{0.5cm}
\fbox{
  \begin{minipage}{8cm} % adjust the width of the box as needed
The results obtained show that the extracted microservices achieve
a median precision score around 0.6. Interestingly, it was observed that the extracted microservices, while not entirely identical to the original microservices, still included most of the classes from the human-built microservices. These results suggest that utilizing machine learning-based approaches for microservice extraction could be a promising direction for software development.
   
  \end{minipage}
}

%%%%%%%%%%%%%%%%%% threats to validity %%%%%%%%%%%%%%%%%%%
\section{Threats to Validity}\label{sec5}
The validity of our findings may be limited in terms of external and internal validity. External validity is limited due to the small sample size of seven diverse applications used in our review. Additionally, the approach has only been tested on Java-based applications at a class level, which may limit generalization to other programming languages and levels of coverage. Regarding internal validity, we need to consider the possibility of coding errors or bugs in our model's implementation, experimental infrastructure and data collection. To address this concern, we took extensive measures, such as conducting code reviews and rigorous testing throughout the development process. Additionally, we ran multiple iterations of experiments to ensure the consistency and reliability of our results.

%%%%%%%%%%%%%%% conclusion %%%%%%%%%%%%%%%

\section{ Conclusion and Future Work}\label{sec6}

In this study, our primary focus was to address the challenge of decomposing a monolithic application into a recommended set of new microservices through a clustering task. The proposed method involved utilizing static analysis tools on the monolithic application's source code to extract class dependencies and encode both structural and semantic information. Subsequently, we employed the adapted-Boosted Mean Shift Clustering algorithm to extract microservices from this encoding, leveraging its advantages, including the ability to infer the number of microservices and robustness to outlier classes.
In addition to conducting a sensitivity analysis on hyperparameters, we evaluated the performance of our approach by comparing it with six baseline methods and assessing the quality of the extracted microservices. Our approach yielded encouraging results across most of the comparison metrics, demonstrating its effectiveness in addressing the decomposition challenge.

To take the granularity level to an advanced level, the approach could be further developed to use methods or functions of the monolith as a basis for decomposition rather than classes. Furthermore, since static analysis does not provide all of the information necessary for a clear understanding of functionalities and interactions during application execution, a hybrid solution incorporating dynamic analysis of the source code could be developed.

%%%%%%%%%%%%%%%%%%%%%%%%%%%%%%%%%%%%%%%%%%%%%%%%%%
%\begin{flusend}
\bibliographystyle{ieeetr}
\bibliography{references}
%\end{flusend}
\end{document}